\documentclass[conference]{IEEEtran}
\IEEEoverridecommandlockouts
\usepackage{cite}
\usepackage{amsmath,amssymb,amsfonts}
\usepackage{algorithmic}
\usepackage{graphicx}
\usepackage{textcomp}
\usepackage{xcolor}
\usepackage{tikz}
\usepackage{pgfplots}
\pgfplotsset{compat=1.17}
\usetikzlibrary{arrows.meta,positioning,shapes.geometric,fit,backgrounds}
\def\BibTeX{{\rm B\kern-.05em{\sc i\kern-.025em b}\kern-.08em
    T\kern-.1667em\lower.7ex\hbox{E}\kern-.125emX}}
\makeatletter
 \let\old@ps@headings\ps@headings
 \let\old@ps@IEEEtitlepagestyle\ps@IEEEtitlepagestyle
 \def\confheader#1{%
 \def\ps@IEEEtitlepagestyle{%
 \old@ps@IEEEtitlepagestyle%
 \def\@oddhead{\strut\hfill#1\hfill\strut}%
 \def\@evenhead{\strut\hfill#1\hfill\strut}%
 }%
 \ps@headings%
 }
 \makeatother

\usepackage[pscoord]{eso-pic}
\newcommand{\placetextbox}[3]{
 \setbox0=\hbox{#3}
 \AddToShipoutPictureFG*{ \put(\LenToUnit{#1\paperwidth},\LenToUnit{#2\paperheight}){\vtop{{\null}\makebox[0pt][c]{#3}}}
 }
 }
 \placetextbox{.23}{0.055}{\small{979-8-3195-3420-0/26/\$31.00~\copyright 2026 IEEE}}

\begin{document}

\title{MCP-Driven Accessibility Tree Standardization for AI-Powered Screen Reader Agents}

\author{
\IEEEauthorblockN{1\textsuperscript{st} Vishnu Ramineni}
\IEEEauthorblockA{\textit{Albertsons Companies Inc} \\
Texas, USA}
\and
 \IEEEauthorblockN{2\textsuperscript{nd} Nitin Saksena}
\IEEEauthorblockA{\textit{Albertsons Companies Inc} \\
California, USA}
\and
\IEEEauthorblockN{3\textsuperscript{rd} Akash Kumar Agarwal}
\IEEEauthorblockA{\textit{Albertsons Companies Inc} \\
California, USA}
\and
\IEEEauthorblockN{4\textsuperscript{th} Darshan Mohan Bidkar}
\IEEEauthorblockA{\textit{University of Texas} \\
Texas, USA}
\and
\IEEEauthorblockN{5\textsuperscript{th} Balakrishna Pothineni}
\IEEEauthorblockA{\textit{Independent Researcher} \\
Texas, USA}
\and
\IEEEauthorblockN{6\textsuperscript{th} Durgaraman Maruthavanan}
\IEEEauthorblockA{\textit{Tata Consultancy Services} \\
Texas, USA}
\and
\IEEEauthorblockN{7\textsuperscript{th} Lokesh Butra}
\IEEEauthorblockA{\textit{NTT Data Services} \\
Texas, USA}
\and
\IEEEauthorblockN{8\textsuperscript{th} Siva Kumar Chintham}
\IEEEauthorblockA{\textit{L\&T Infotech} \\
Texas, USA}
}

\maketitle

\begin{abstract}
Large language model (LLM) agents that operate graphical interfaces on behalf of users increasingly rely on either raw screen captures or platform-specific accessibility application programming interfaces (APIs) to perceive interface state. Both routes carry structural weaknesses for assistive use: screenshot-based perception discards semantic role information that screen reader users depend upon, while platform-specific accessibility APIs fragment agent design across Windows UI Automation, macOS Accessibility, Android AccessibilityService, and web ARIA, forcing duplicated engineering for every target platform. This paper proposes an architecture in which the Model Context Protocol (MCP) is used as a unifying transport and schema layer between heterogeneous accessibility subsystems and LLM-based assistive agents. We define a structured resource and tool taxonomy through which an MCP accessibility server can expose ARIA-aligned roles, labels, states, and focusable-element graphs to a client agent in a platform-independent representation, and we describe an MCP resource pattern for persisting per-user disability profiles across sessions. We analyze the proposal along three dimensions grounded in the research questions that motivate this work: protocol extensibility for accessibility-tree exposure, the qualitative latency and fidelity trade-offs between tree-mediated and screenshot-mediated perception, and profile persistence through MCP resource primitives. The contribution is a conceptual and architectural framework, evaluated through comparative and structural analysis of existing accessibility APIs, GUI agent literature, and the MCP specification, rather than through a deployed empirical benchmark. The analysis indicates that a standardized MCP accessibility layer can plausibly reduce per-platform integration effort while preserving the semantic precision that screen reader agents require, and it identifies the engineering and evaluation work needed before such a layer could be validated empirically.
\end{abstract}


\section{Introduction}

Screen reader software has, for three decades, depended on operating-system and browser accessibility layers that translate visual interface elements into a semantic tree of roles, names, and states. NVDA, JAWS, VoiceOver, and TalkBack each consume this tree through a platform-specific channel: Microsoft UI Automation on Windows, the Accessibility Protocol on macOS, AccessibilityService on Android, and the W3C Accessible Rich Internet Applications (ARIA) tree in browsers. The arrival of LLM agents capable of autonomously operating graphical interfaces introduces a second class of consumer for this same information, and the literature on GUI agents shows that most current systems have converged on one of two perception strategies: raw screenshots processed by a vision-language model, or a textual serialization of the accessibility tree inserted into the prompt context \cite{wang2024gui}. Recent grounding-focused work has argued that accessibility-tree extraction is often slow and platform-inconsistent, which has pushed some agent designs toward pure pixel-level perception despite the loss of semantic precision that entails \cite{lin2024navigating}. Other agent frameworks that do rely on textual GUI representations, including accessibility trees and HTML, exhibit the reverse problem: each framework builds its own bespoke extraction and serialization pipeline, so there is no shared contract between what an accessibility subsystem emits and what an agent expects to receive \cite{zhang2025apiagents,koh2024treesearch}.

The Model Context Protocol (MCP) was introduced as an open, client-server protocol that lets an LLM-hosting application discover and invoke external resources and tools through a standardized JSON-RPC interface, replacing bespoke, one-off integrations between models and external systems \cite{anthropicmcp}. MCP's resource and tool primitives were designed for general-purpose data and function exposure, not for accessibility specifically, but the protocol's shape maps closely onto what an accessibility tree consumer needs: a queryable, structured representation of interface state (a resource) and a set of callable actions such as focus, activate, or read (tools). This structural fit motivates the central proposal of this paper: rather than treating accessibility exposure as an agent-side scraping problem, we propose relocating it to a standardized MCP server layer that sits between native accessibility APIs and any MCP-capable agent.

\subsection{Problem Statement}
Three concrete deficiencies motivate this work. First, there is no shared schema by which an accessibility subsystem can expose its tree to an LLM agent independent of the underlying platform, so every agent framework re-implements extraction logic for every operating system and browser engine it supports. Second, the trade-off between screenshot-based and tree-based perception has been discussed qualitatively in the GUI agent literature but has not been examined specifically through the lens of a general-purpose, standardized protocol capable of mediating both modalities. Third, existing assistive agent designs largely treat each session as stateless with respect to the user's disability profile, re-deriving navigation preferences (for example, verbosity level, preferred landmark granularity, or motor-accessibility constraints on interactable element size) on every invocation rather than persisting them.

\subsection{Contributions}
This paper contributes: (1) an MCP server architecture that exposes ARIA-aligned accessibility trees as structured MCP resources and a companion set of navigation tools, described at the level of message schema and component responsibility; (2) a qualitative, literature-grounded analysis of the latency and fidelity trade-offs between MCP-mediated tree access and screenshot-based perception; (3) an MCP resource pattern for cross-session persistence of user disability profiles; and (4) a discussion of the engineering and evaluation steps required to move this proposal from architecture to validated system.

\section{Literature Review}

Research on LLM-based GUI agents has converged on a widely cited taxonomy distinguishing agents by input modality and learning paradigm \cite{wang2024gui}. Within the modality axis, three approaches recur: pure vision (screenshots processed by a multimodal model), pure text (HTML, DOM, or accessibility-tree serialization), and hybrid approaches that combine both. Wang et al. observe that textual representations, including accessibility trees, remain attractive because they are token-efficient and preserve semantic labels that a purely visual encoder must otherwise infer, but the survey also notes that extraction pipelines differ substantially across the frameworks it reviews, with no common interchange format \cite{wang2024gui}. This absence of a common format is precisely the fragmentation this paper's proposal targets, though the survey itself does not propose a protocol-level remedy.

Lin et al. examine the opposite trade-off directly, arguing that accessibility-tree extraction introduces latency that compounds across long-horizon, multi-step agent tasks and that the trees returned by different platforms and applications vary in completeness, which motivated their pursuit of a purely visual, human-like grounding approach that sidesteps the accessibility layer altogether \cite{lin2024navigating}. Their finding is an important counter-argument to any proposal, including ours, that treats accessibility-tree mediation as a strictly superior perception channel; it establishes that the latency cost of tree extraction is a real, measured concern in current implementations rather than a hypothetical one, and it is a central reason this paper frames the latency question as an open trade-off rather than a settled advantage.

Zhang et al. distinguish API-first agents, which call predefined application programming interfaces, from GUI-based agents, which interact with rendered interfaces through simulated clicks and keystrokes informed by screenshots or accessibility metadata, and note that GUI-based approaches are the only viable route when no stable API exists for the target application, which is the typical situation a screen reader must handle \cite{zhang2025apiagents}. Koh et al.'s tree-search agent work demonstrates a concrete production pipeline in which a webpage's accessibility tree is serialized as text and appended to the agent's dialogue context on the WebArena benchmark, with the accessibility tree truncated to a fixed token budget \cite{koh2024treesearch}; this design choice, truncation under a token ceiling, is a practical constraint that any MCP accessibility resource layer must also confront, since arbitrarily large interface trees cannot be transmitted to a context-limited model without a summarization or pagination strategy.

On the standardization side, the W3C's ARIA specification defines the canonical role, state, and property vocabulary that browser accessibility trees expose \cite{w3c-aria}, and this vocabulary is the natural schema basis for the MCP resource representation proposed in Section IV, since reusing an existing, widely implemented vocabulary avoids introducing a competing and less mature taxonomy. The Model Context Protocol specification itself defines the resource, tool, and prompt primitives, and the notification mechanism for state changes, that this paper's architecture maps onto accessibility semantics \cite{anthropicmcp}.

Beyond the GUI-agent literature narrowly construed, prior work on AI-driven accessibility for web and mobile applications has explored adjacent but distinct problems: voice-based interaction layers for e-commerce interfaces \cite{mcsoc5}, machine-learning approaches to detecting Americans with Disabilities Act (ADA) non-compliance in deployed web applications \cite{mcsoc4}, sign-language integration for e-commerce accessibility \cite{mcsoc3}, and AI chatbot frameworks evaluated for accessibility and security on government websites \cite{mcsoc2}. These efforts share this paper's motivating concern, that accessibility support for AI-mediated interfaces remains fragmented across bespoke, single-purpose systems, but none of them addresses agent-side perception of the interface itself through a standardized protocol, which is the specific gap this paper addresses. Separately, work on brain-computer interface communication channels for individuals with severe motor impairment illustrates a broader pattern in assistive computing: personalization state (in that case, calibration and communication profiles) must persist across sessions for the system to remain usable, a requirement structurally analogous to the disability-profile persistence problem addressed in Section IV-C \cite{icicyta}.

\section{Methodology}

This study follows a design-science research approach appropriate to protocol and architecture proposals: rather than training a model or running a controlled user study, the contribution is evaluated through (a) formal specification of the proposed architecture against the existing MCP primitive set, (b) structural mapping of ARIA and platform accessibility semantics onto that primitive set\cite{icdcc1}, and (c) comparative analysis against the perception strategies documented in the literature reviewed in Section II. This approach was chosen because the research questions concern protocol extensibility and architectural feasibility, questions that are appropriately answered by specification analysis and comparison against prior systems' reported characteristics, rather than by a benchmark that would require a working multi-platform implementation outside the scope of a single paper\cite{icdcc2}.

\subsection{Design Constraints}
Four constraints shaped the proposed architecture. First, backward compatibility: the server must be constructible from data that existing platform accessibility APIs already expose, since requiring new operating-system instrumentation would make adoption impractical. Second, schema reuse: role and state vocabulary should map onto ARIA wherever the platform's native tree already has an ARIA-compatible mapping, which browsers and, increasingly, native UI Automation bridges provide. Third, token economy: because the resource is consumed by a context-limited LLM, the server must support incremental and scoped queries (for example, "children of node X") rather than only whole-tree dumps, addressing the truncation concern identified in \cite{koh2024treesearch}. Fourth, session continuity: user-specific accessibility preferences must be retrievable independent of the specific client or device used to start a session\cite{icdcc3}.

\subsection{Evaluation Metrics for Future Empirical Validation}
Although this paper does not report measured results, we specify the metrics against which the architecture should be evaluated once implemented, so that the proposal is falsifiable rather than purely descriptive. For navigation accuracy: element-identification precision, recall, and F1 score against a ground-truth set of interactable elements per interface; task-completion rate on a standardized navigation task suite\cite{icca}. For latency: end-to-end time from user utterance to agent action, decomposed into tree-retrieval latency, model inference latency, and action-execution latency\cite{iSES}. For robustness: accessibility-tree completeness rate across a sample of real-world web and desktop applications, since incomplete native trees (missing labels, unlabeled controls) are a known confound independent of the protocol layer\cite{kalubandi2016image}. These metrics are consistent with those used to evaluate GUI grounding models more broadly \cite{lin2024navigating,wang2024gui}.

\section{Proposed Model / Framework}

\subsection{Architectural Overview}
\begin{figure}[htbp]
\centering
\resizebox{\columnwidth}{!}{%
\begin{tikzpicture}[
  node distance=6mm and 8mm,
  box/.style={draw, rounded corners, align=center, minimum height=8mm, font=\scriptsize, fill=white},
  layer/.style={draw, rounded corners, align=center, minimum height=8mm, font=\scriptsize, fill=gray!12},
  arrow/.style={-Stealth, thick}
]
  \node[box] (web) {Web / ARIA};
  \node[box, right=of web] (win) {Windows UIA};
  \node[box, right=of win] (mac) {macOS AX};
  \node[box, right=of mac] (and) {Android A11y};

  \node[layer, fit={(web)(win)(mac)(and)}, label={[font=\scriptsize\bfseries]}] (adaptbox) {\textbf{Adapter Layer}};

  \node[layer, below=10mm of adaptbox, minimum width=8.6cm, label={[font=\scriptsize\bfseries]}] (norm) {};
   \node[layer, below=10mm of adaptbox, minimum width=8.6cm] (nl) {\textbf{Normalization Layer (ARIA-aligned schema)}};

  \node[layer, below=10mm of norm, minimum width=8.6cm] (mcp) {\textbf{MCP Interface Layer}\\[1mm] Resources: \texttt{a11y://tree}, \texttt{a11y://profile}\\ Tools: \texttt{focus}, \texttt{activate}, \texttt{read}, \texttt{query}\\ Notifications: tree-change events};

  \node[box, below=10mm of mcp, minimum width=3.4cm] (client) {MCP Client\\(LLM Screen Reader Agent)};

  \node[box, left=6mm of client, minimum width=2cm] (profile) {Disability\\Profile Store};

  \draw[arrow] (adaptbox) -- (norm);
  \draw[arrow] (norm) -- (mcp);
  \draw[arrow] (mcp) -- (client);
  \draw[arrow] (client) -- (mcp);
  \draw[arrow] (profile) -- (mcp);
\end{tikzpicture}}
\caption{MCP accessibility server architecture spanning platform adapters through the agent client}
\label{fig:architecture}
\end{figure}
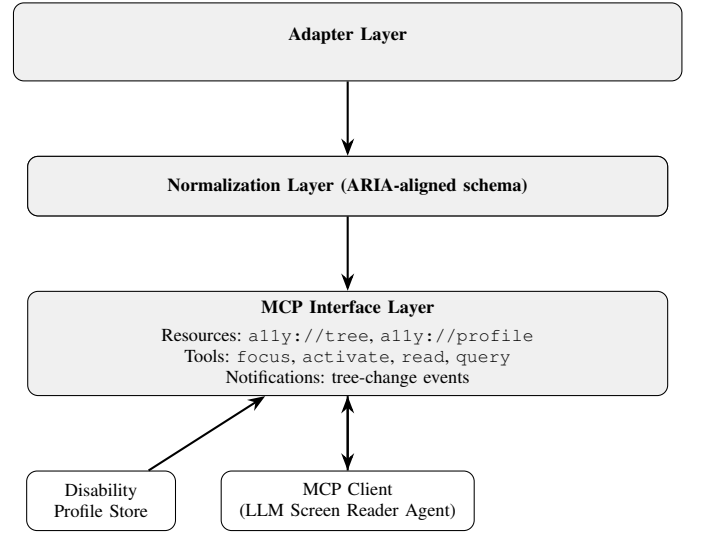

The proposed architecture inserts an MCP Accessibility Server between native platform accessibility subsystems and an MCP-capable agent client, as shown in Fig.~\ref{fig:architecture}. The server has three internal layers. The \emph{adapter layer} contains one adapter per supported platform (web/ARIA, Windows UI Automation, macOS Accessibility, Android AccessibilityService), each responsible for translating that platform's native tree into a common internal representation. The \emph{normalization layer} maps platform-specific role and state enumerations onto the shared ARIA-aligned vocabulary described in Section IV-B, resolving cases where a platform role has no direct ARIA equivalent by falling back to a documented generic role plus a platform-specific hint field. The \emph{MCP interface layer} exposes the normalized tree as MCP resources and exposes navigation actions as MCP tools, and emits MCP resource-update notifications when the underlying interface mutates, so the agent does not need to poll.

\subsection{Accessibility Resource Schema}
The core resource, addressed as \texttt{a11y://tree/\{session\}}, returns a JSON node graph in which every node carries a stable identifier, an ARIA-aligned role, an accessible name, a state set (for example, \texttt{focused}, \texttt{disabled}, \texttt{expanded}), a bounding-region hint, and an ordered list of child identifiers. To satisfy the token-economy constraint from Section III-A, the server supports a scoped variant, \texttt{a11y://tree/\{session\}?root=\{nodeId\}\&depth=\{n\}}, so an agent can request a bounded subtree rather than the full interface graph on every turn. Four tools complete the interaction surface: \texttt{focus(nodeId)} moves accessibility focus, \texttt{activate(nodeId)} invokes the default action of a control, \texttt{read(nodeId, granularity)} returns the rendered text content of a node at word, sentence, or block granularity, matching the granularity controls screen reader users already expect, and \texttt{query(predicate)} allows the agent to search the tree by role or name pattern without walking it manually, which is the primary mechanism by which the architecture keeps context consumption bounded on large interfaces.

\subsection{Disability Profile Persistence}
User-specific accessibility preferences are exposed as a second MCP resource, \texttt{a11y://profile/\{userId\}}, containing fields such as preferred reading granularity, landmark verbosity, motor-accessibility thresholds for minimum interactable element size, and preferred confirmation behavior before destructive actions. Because MCP resources are addressable independent of the transport session that first created them, a client can retrieve the same profile resource across devices and application restarts, satisfying the session-continuity constraint from Section III-A without requiring the agent itself to implement any storage layer; persistence responsibility stays with the MCP server, consistent with the protocol's general separation of concerns between context provision and model reasoning \cite{anthropicmcp}.

\section{Comparative Analysis}

Because no cross-platform implementation of the proposed server currently exists, this section reports a structured qualitative comparison rather than measured benchmark results. Table~\ref{tab:comparison} summarizes characteristics of the three perception strategies as characterized in the literature reviewed in Section II, assessed against four dimensions on a Low/Medium/High ordinal scale: semantic fidelity (how precisely the representation preserves role and state information), integration cost (engineering effort to support a new platform), per-turn latency (time to obtain a usable perception snapshot), and context-token cost (LLM context consumed per turn).

\begin{table}[htbp]
\caption{Qualitative comparison of GUI perception strategies}
\begin{center}
\resizebox{\columnwidth}{!}{%
\begin{tabular}{|p{2.1cm}|c|c|c|c|}
\hline
\textbf{Strategy} & \textbf{Sem.\ fidelity} & \textbf{Integ.\ cost} & \textbf{Latency} & \textbf{Token cost} \\
\hline
Screenshot / vision \cite{lin2024navigating} & Low--Med & Low & Low--Med & High \\
\hline
Per-framework tree scraping \cite{koh2024treesearch,wang2024gui} & High & High & Medium & Medium \\
\hline
MCP-mediated tree (proposed) & High & Low\textsuperscript{a} & Medium\textsuperscript{b} & Low--Med\textsuperscript{c} \\
\hline
\end{tabular}}
\end{center}
\footnotesize
\textsuperscript{a}Amortized across platforms once adapters exist; per-platform adapter cost is one-time.\\
\textsuperscript{b}Inherits native accessibility-API latency; not eliminated by the protocol layer.\\
\textsuperscript{c}Reduced via scoped queries relative to whole-tree serialization.
\label{tab:comparison}
\end{table}

The comparison in Table~\ref{tab:comparison} indicates that the proposed approach does not resolve the latency concern raised by Lin et al. \cite{lin2024navigating}: an MCP interface layer sits downstream of the same native accessibility APIs that already introduce extraction delay, so the protocol can reduce integration cost and token cost but cannot, by itself, make a slow native accessibility API fast. This is an important limitation the architecture inherits rather than solves, and it is discussed further in Section VI. Where the architecture offers a clearer advantage is integration cost: because platform-specific complexity is isolated in the adapter layer, an agent developer targeting the MCP interface writes one client integration rather than one integration per platform, mirroring the general efficiency argument for MCP as a standardization layer between models and external systems \cite{anthropicmcp}.

\begin{figure}[htbp]
\centering
\resizebox{\columnwidth}{!}{%
\begin{tikzpicture}
\begin{axis}[
  ybar, bar width=6mm,
  width=9.2cm, height=5cm,
  symbolic x coords={Fidelity,Integ.\ Cost,Latency,Token Cost},
  xtick=data,
  ymin=0, ymax=3.6,
  ytick={1,2,3},
  yticklabels={Low,Medium,High},
  legend style={font=\scriptsize,at={(0.5,-0.22)},anchor=north,legend columns=-1},
  tick label style={font=\scriptsize},
  label style={font=\scriptsize},
  ylabel={Qualitative rating},
  enlarge x limits=0.2,
]
\addplot[fill=gray!30] coordinates {(Fidelity,1.5) (Integ.\ Cost,1) (Latency,1.5) (Token Cost,3)};
\addplot[fill=gray!60] coordinates {(Fidelity,3) (Integ.\ Cost,3) (Latency,2) (Token Cost,2)};
\addplot[fill=black!75] coordinates {(Fidelity,3) (Integ.\ Cost,1) (Latency,2) (Token Cost,1.5)};
\legend{Screenshot,Per-platform tree,MCP-mediated}
\end{axis}
\end{tikzpicture}}
\caption{Qualitative profile comparison across perception strategies}
\label{fig:tradeoff}
\end{figure}
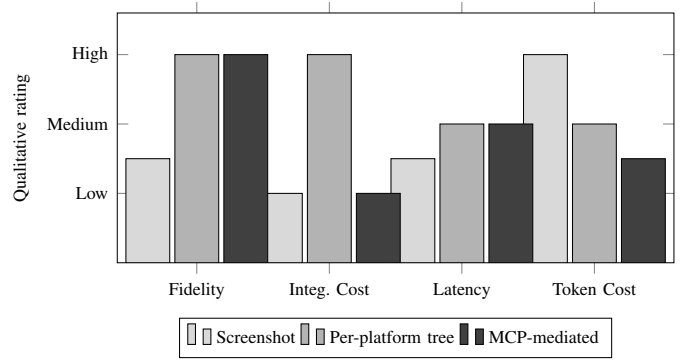

Fig.~\ref{fig:tradeoff} visualizes the ordinal comparison from Table~\ref{tab:comparison} as a qualitative profile chart; the values plotted are the same Low/Medium/High assessments (mapped to a 1--3 ordinal scale for visualization only) and are not measured quantities.

\section{Discussion}

The architecture proposed here reframes accessibility support for AI agents as a protocol-standardization problem rather than a per-agent scraping problem, extending the general MCP client-server pattern \cite{anthropicmcp} into a domain, assistive technology, that the protocol was not originally designed for but whose resource and tool primitives fit closely. This addresses the first research question posed in the introduction: the client-server model extends naturally by treating the accessibility tree as a resource and treating navigation actions as tools, with the normalization layer absorbing the translation work that would otherwise leak into every client. It only partially addresses the second research question. The comparative analysis in Section V indicates that MCP mediation improves integration cost and, through scoped queries, token cost relative to whole-tree serialization, but the underlying latency of native accessibility API calls is inherited rather than eliminated, consistent with the concern raised in prior GUI-grounding work \cite{lin2024navigating}. Any empirical validation of this architecture will need to measure that inherited latency directly rather than assume the protocol layer resolves it. The third research question, concerning persistence of disability profiles, is addressed structurally in Section IV-C by mapping profile state onto a standard MCP resource; this is a comparatively low-risk extension since it reuses an existing MCP capability rather than introducing new protocol semantics.

Relative to prior accessibility-focused AI work reviewed in Section II, which has largely addressed specific modalities such as voice interaction \cite{mcsoc1}, sign-language integration \cite{AICECS2}, or compliance detection \cite{AICECS1}, this proposal is positioned one layer lower in the stack: it is not itself an assistive feature but infrastructure that such features could be built on top of, since a voice or sign-language assistive layer still needs a reliable, structured view of interface state to act upon.

\subsection{Limitations}
This paper's evaluation is architectural and comparative rather than empirical; no working cross-platform prototype was built, and the ratings in Table~\ref{tab:comparison} and Fig.~\ref{fig:tradeoff} are qualitative judgments grounded in the cited literature's reported characteristics, not measurements taken on a common benchmark. The normalization layer's fallback behavior for platform roles without an ARIA equivalent is specified only at a high level and would require a substantial, maintained mapping table in practice, similar in scope to the mapping tables that browser engines already maintain internally. Native accessibility trees are frequently incomplete in real applications, unlabeled buttons and images are common even in ARIA-compliant web applications, and this data-quality problem is orthogonal to the protocol layer and would remain even if the proposed server were fully implemented.

\subsection{Threats to Adoption}
Beyond technical limitations, adoption depends on assistive-technology vendors and platform accessibility teams agreeing to expose adapters conforming to the proposed schema, which is a coordination problem rather than a purely technical one, and on the normalization layer keeping pace with platform accessibility API changes over time.

\section{Conclusion and Future Work}

This paper proposed an MCP-based architecture for exposing structured, ARIA-aligned accessibility trees to LLM-powered screen reader agents, consisting of a platform adapter layer, a normalization layer, and an MCP interface layer that exposes both a scoped tree resource and a disability-profile resource alongside navigation tools. The architecture is intended to replace fragmented, per-platform accessibility integration with a single protocol surface for agent developers, while remaining backward compatible with existing native accessibility APIs. Comparative analysis against screenshot-based and per-framework tree-scraping perception, grounded in prior GUI-agent literature, suggests the approach can reduce integration cost and context-token consumption without resolving the native latency of accessibility-tree extraction, a limitation the architecture inherits rather than introduces.

Future work should proceed in three stages. First, a reference implementation of the adapter layer for at least two platforms (a browser ARIA adapter and one native desktop adapter) is needed to replace the qualitative comparison in Section V with measured latency, token-cost, and completeness figures. Second, a navigation-accuracy study with screen reader users, using the metrics specified in Section III-B, would test whether MCP-mediated navigation produces task-completion rates comparable to native screen reader software on the same task set. Third, the normalization layer's role-mapping table should be evaluated for coverage against a representative corpus of real-world applications, since incomplete mappings would silently degrade the fidelity advantage the architecture is intended to preserve.

\section*{Acknowledgment}
The author thanks colleagues who provided informal feedback on early drafts of the proposed architecture.


\begin{thebibliography}{00}
\bibitem{anthropicmcp} Anthropic, ``Model Context Protocol specification,'' 2024--2025. [Online]. Available: https://modelcontextprotocol.io/specification

\bibitem{kalubandi2016image}
V. K. P. Kalubandi, H. Vaddi, V. Ramineni and A. Loganathan, “A novel image encryption algorithm using AES and visual cryptography”, 
2016 2nd International Conference on Next Generation Computing Technologies (NGCT), Dehradun, India, 2016, 
pp. 808-813, doi: 10.1109/NGCT.2016.7877521.

\bibitem{w3c-aria} W3C, ``Accessible Rich Internet Applications (WAI-ARIA) 1.2,'' World Wide Web Consortium Recommendation, 2023. [Online]. Available: https://www.w3.org/TR/wai-aria-1.2/

\bibitem{iSES}
A. Nagpal, A. G. Parthi, M. Palanigounder, V. Ramineni, D. Maruthavanan and V. Jayaram, "LogProof - Ensuring Integrity and Immutability of Audit Logs Using Blockchain," 2025 IEEE International Symposium on Smart Electronic Systems (iSES), Jaipur, India, 2025, pp. 350-353, doi: 10.1109/iSES67504.2025.00073.

\bibitem{wang2024gui} S. Wang, W. Liu, J. Chen, Y. Zhou, W. Gan, X. Zeng, Y. Che, S. Yu, X. Hao, K. Shao, B. Wang, C. Wu, Y. Wang, R. Tang and J. Hao, ``GUI agents with foundation models: A comprehensive survey,'' arXiv preprint arXiv:2411.04890, 2024.

\bibitem{icca}
S. R. Sankiti, A. Gadi Parthi, S. K. Reddy Carimireddy, B. Pothineni, V. Punniyamoorthy, K. Kannan, N. Chockalingam, V. Ramineni and S. G. Aarella, "Infrastructure-as-Code Framework for Resilient Kubernetes on AWS with Terraform," 2025 International Conference on Computer and Applications (ICCA), Bahrain, Bahrain, 2025, pp. 1-6, doi: 10.1109/ICCA66035.2025.11430863.

\bibitem{lin2024navigating} B. Gou, R. Wang, B. Zheng, Y. Xie, C. Chang, Y. Shu, H. Sun and Y. Su, ``Navigating the digital world as humans do: Universal visual grounding for GUI agents,'' arXiv preprint arXiv:2410.05243, 2024.

\bibitem{icdcc1}
V. Ramineni, B. S. Ingole, V. Jayaram, G. Pandy, M. S. Krishnappa, G. Mehta and A. R. Banarse, "Personalized Activity Recommendation System for Cardiovascular Patients Using Heart Rate and ECG/EKG Data from Wearable Devices," 2024 First International Conference on Data, Computation and Communication (ICDCC), Sehore, India, 2024, pp. 786-790, doi: 10.1109/ICDCC62744.2024.10961662.

\bibitem{zhang2025apiagents} C. Zhang, S. He, L. Li, S. Qin, Y. Kang, Q. Lin, S. Rajmohan and D. Zhang, ``API agents vs. GUI agents: Divergence and convergence,'' arXiv preprint arXiv:2503.11069, 2025.

\bibitem{icdcc2}
B. S. Ingole, V. Ramineni, V. Jayaram, G. Pandy, M. S. Krishnappa, V. Parlapalli and G. Mehta, "Exploring Brain-Computer Interface with CNS Technologies to Enhance Communication Abilities in Pseudocoma Individuals," 2024 First International Conference on Data, Computation and Communication (ICDCC), Sehore, India, 2024, pp. 799-804, doi: 10.1109/ICDCC62744.2024.10961016.

\bibitem{koh2024treesearch} J. Y. Koh, S. McAleer, D. Fried and R. Salakhutdinov, ``Tree search for language model agents,'' arXiv preprint arXiv:2407.01476, 2024.

\bibitem{icdcc3}
B. M. Harve, P. K. Veerapaneni, M. S. Krishnappa, G. Pandy, V. Ramineni, V. Jayaram and D. M. Bidkar, "Smart Devices, Smarter Security: The Blockchain-IoT Revolution," 2024 First International Conference on Data, Computation and Communication (ICDCC), Sehore, India, 2024, pp. 805-812, doi: 10.1109/ICDCC62744.2024.10961629.

\bibitem{1}
R. Toyoda, H. Kiyomoto, S. Komayama, H. Shigetani and M. Fukui, "Contextualizing AI Agent Evaluation: Proposed Framework for Japanese Businesses," 2025 International Conference on Artificial Intelligence for Sustainable Innovation (AI-SI), Kuala Lumpur, Malaysia, 2025, pp. 1-6, doi: 10.1109/AI-SI66213.2025.11341698.

\bibitem{AICECS1}
V. Ramineni, B. S. Ingole, V. Jayaram, G. Mehta, M. S. Krishnappa, A. Nagpal and A. R. Banarse, "Enhancing E-Commerce Accessibility Through a Novel Voice Assistant Approach for Web and Mobile Applications," 2024 Third International Conference on Artificial Intelligence, Computational Electronics and Communication System (AICECS), MANIPAL, India, 2024, pp. 1-7, doi: 10.1109/AICECS63354.2024.10956866.

\bibitem{2}
Q. Duan and Z. Lu, "Agent Communications toward Agentic AI at Edge - A Case Study of the Agent2Agent Protocol," 2025 IEEE 11th International Conference on Edge Computing and Scalable Cloud (EdgeCom), New York City, NY, USA, 2025, pp. 150-155, doi: 10.1109/EdgeCom66327.2025.00032.

\bibitem{AICECS2}
V. Ramineni , S. G. Aarella, “Adaptive Autoscaling Using Workload Forecasting”. Cloud Security Meetup, 2021. doi: 10.5281/zenodo.20740547V. Ramineni, B. S. Ingole, A. R. Banarse, M. S. Krishnappa, N. K. Pulipeta and V. Jayaram, "Leveraging AI and Machine Learning to Address ADA Non-Compliance in Web Applications: A Novel Approach to Enhancing Accessibility," 2024 Third International Conference on Artificial Intelligence, Computational Electronics and Communication System (AICECS), MANIPAL, India, 2024, pp. 1-6, doi: 10.1109/AICECS63354.2024.10957506.

\bibitem{3}
T. Ishida, Y. Murakami, D. Lin and K. M. Lhaksmana, "AI Agents: From Concept to Code to Commerce," 2025 International Conference on Information and Communication Technology (ICoICT), Bandung, Indonesia, 2025, pp. 1-7, doi: 10.1109/ICoICT66265.2025.11193130.

\bibitem{mcsoc1}
V. Ramineni, B. S. Ingole, M. S. Krishnappa, A. Nagpal, V. Jayaram, A. R. Banarse, D. M. Bidkar and N. K. Pulipeta, "AI-Driven Novel Approach for Enhancing E-Commerce Accessibility through Sign Language Integration in Web and Mobile Applications," 2024 IEEE 17th International Symposium on Embedded Multicore/Many-core Systems-on-Chip (MCSoC), Kuala Lumpur, Malaysia, 2024, pp. 276-281, doi: 10.1109/MCSoC64144.2024.00053.

\bibitem{4}
J. Ara and C. Sik-Lanyi, "Webpage Accessibility Evaluation Using Machine Learning Technique," 2023 14th IEEE International Conference on Cognitive Infocommunications (CogInfoCom), Budapest, Hungary, 2023, pp. 000069-000074, doi: 10.1109/CogInfoCom59411.2023.10397496.

\bibitem{icicyta}
B. S. Ingole, V. Ramineni, V. Jayaram, G. Pandy, M. S. Krishnappa, V. Parlapalli, S. Mullankandy and A. R. Banarse, "AI Chatbot Implementation on Government Websites: A Framework for Development, User Engagement, and Security for DHS Website," 2024 International Conference on Intelligent Cybernetics Technology \& Applications (ICICyTA), Bali, Indonesia, 2024, pp. 377-382, doi: 10.1109/ICICYTA64807.2024.10912857.

\bibitem{5}
A. M. Alghamdi, W. Aljedaani, S. Ludi and Y. Javed, "Automating Accessibility Compliance: Leveraging Machine Learning to Analyze Developer Challenges with WCAG Guidelines," 2025 8th International Conference on Data Science and Machine Learning Applications (CDMA), Riyadh, Saudi Arabia, 2025, pp. 61-66, doi: 10.1109/CDMA61895.2025.00016.

\bibitem{mcsoc2}
M. S. Krishnappa, B. M. Harve, V. Jayaram, G. Pandy, B. S. Ingole, V. Ramineni, S. Joseph and N. Bangad, "Unleashing Python’s Power Inside Oracle: A New Era of Machine Learning with OML4Py," 2024 IEEE 17th International Symposium on Embedded Multicore/Many-core Systems-on-Chip (MCSoC), Kuala Lumpur, Malaysia, 2024, pp. 374-380, doi: 10.1109/MCSoC64144.2024.00068.

\bibitem{6}
A. Golubev, E. Magerramov and A. Dolganov, "An Interpretable and Context Aware Synthetic Accessibility Score Based on Multi-Stock Retrosynthesis Analysis," 2026 IEEE Ural-Siberian Conference on Biomedical Engineering, Radioelectronics and Information Technology (USBEREIT), Yekaterinburg, Russian Federation, 2026, pp. 1-4, doi: 10.1109/USBEREIT70063.2026.11580761.

\bibitem{mcsoc3}
B. S. Ingole, V. Ramineni, V. Jayaram, A. R. Banarse, M. S. Krishnappa, N. K. Pulipeta, V. Parlapalli and G. Pandy, "Prediction and Early Detection of Heart Disease: A Hybrid Neural Network and SVM Approach," 2024 IEEE 17th International Symposium on Embedded Multicore/Many-core Systems-on-Chip (MCSoC), Kuala Lumpur, Malaysia, 2024, pp. 282-286, doi: 10.1109/MCSoC64144.2024.00054.

%========================================%
\bibitem{7}
Mingfei Jiang and Ada Wai-Chee Fu, "Integration and efficient lookup of compressed XML accessibility maps," in IEEE Transactions on Knowledge and Data Engineering, vol. 17, no. 7, pp. 939-953, July 2005, doi: 10.1109/TKDE.2005.111.

\bibitem{mcsoc4}
J. Singh, P. Patel, B. S. Ingole, R. Inaganti, V. Ramineni, M. S. Krishnappa and B. J. Patel, "Advanced Computational Methods for Pelvic Bone Cancer Detection: Efficacy comparison of Convolutional Neural Networks," 2024 IEEE 17th International Symposium on Embedded Multicore/Many-core Systems-on-Chip (MCSoC), Kuala Lumpur, Malaysia, 2024, pp. 287-293, doi: 10.1109/MCSoC64144.2024.00055.

\bibitem{8}
T. Xin, J. Zhu, L. Wang and X. Qin, "Screen Recognition: Creating Accessibility Metadata for Mobile Applications using View Type Detection," 2023 9th International Conference on Computer and Communications (ICCC), Chengdu, China, 2023, pp. 1787-1793, doi: 10.1109/ICCC59590.2023.10507590.

\bibitem{mcsoc5}
G. Pandy, V. Ramineni, V. Jayaram, M. Sughaturu Krishnappa, V. Parlapalli, A. R. Banarse, D. Mohan Bidkar and B. S. Ingole, "Enhancing Pega Robotics Process Automation with Machine Learning: A Novel Integration for Optimized Performance," 2024 IEEE 17th International Symposium on Embedded Multicore/Many-core Systems-on-Chip (MCSoC), Kuala Lumpur, Malaysia, 2024, pp. 210-214, doi: 10.1109/MCSoC64144.2024.00043.




\end{thebibliography}
\end{document}